\documentclass[12pt]{article}

\usepackage{times,color, bm}
\usepackage{epsfig,amsmath,amssymb,amsfonts,mathrsfs,braket}
\usepackage{subfigure,graphicx,epstopdf}
\usepackage[colorlinks]{hyperref}
\hypersetup{linkcolor = {black},citecolor = {black},	urlcolor = {black}}
\usepackage{cite}
\usepackage{upgreek}
\usepackage[hyphenbreaks]{breakurl}

\newenvironment{sciabstract}{%
\begin{quote} \bf}
{\end{quote}}

\newcounter{lastnote}

\begin{document}

\title{A Reconfigurable Photonic Processor for NP-complete Problems}
\author{Xiao-Yun Xu$^{1,2}$, Tian-Yu Zhang$^{1,2}$, Zi-Wei Wang$^{1,2}$, Chu-Han Wang$^{1,2}$, Xian-Min Jin$^{1,2,3,4\ast}$ \\
\normalsize{$^1$Center for Integrated Quantum Information Technologies (IQIT), School of Physics and}\\
\normalsize{Astronomy and State Key Laboratory of Advanced Optical Communication Systems} \\
\normalsize{and Networks, Shanghai Jiao Tong University, Shanghai 200240, China}\\
\normalsize{$^2$Hefei National Laboratory, Hefei 230088, China}\\ 
\normalsize{$^3$TuringQ Co., Ltd., Shanghai 200240, China}\\
\normalsize{$^4$Chip Hub for Integrated Photonics Xplore (CHIPX), Shanghai Jiao Tong University, }\\ 
\normalsize{Wuxi 214000, China}\\
\normalsize{$^\ast$E-mail: xianmin.jin@sjtu.edu.cn}
}

\date{}

\baselineskip 24pt

\maketitle

\begin{sciabstract}
NP-complete problems are widely and deeply involved in various real-life scenarios while still intractable to solve efficiently on conventional computers. It is of great practical significance to construct versatile computing architectures that solve NP-complete problems with computational advantage. Here, we present a reconfigurable photonic processor to efficiently solve a benchmark NP-complete problem, the subset sum problem (SSP). We show that in the case of successive primes, the photonic processor has genuinely surpassed commercial electronic processors launched recently by taking advantages of the high propagation speed and vast parallelism of photons and state-of-the-art integrated photonic technology. Moreover,  we are able to program the photonic processor to tackle different problem instances relying on the tunable integrated modules, variable split junctions, which can be used to build a fully reconfigurable architecture potentially allowing $2^{N}$ configurations at most. Our experiments confirm the potential of the photonic processor as a versatile and efficient computing platform, suggesting a possible practical route to solving computationally hard problems at large scale.\\
\end{sciabstract}

\noindent Though integrated circuit technology has experienced a rapid development and greatly enhanced our computing power in the past few decades \cite{Venema2011silicon}, a myriad of computational problems are still hard to efficiently solve \cite{Moore2016, garey2002computers,wei2022trends}. The hardness mostly lies in the huge consumption of resource, especially time resource that is irreversible and non-recyclable \cite{sipser2012introduction}. According to computational complexity theory \cite{garey2002computers,sipser2012introduction}, problems in the class NP-complete are out of the reach of traditional electronic computers, which are generally regarded as physical embodiments of deterministic Turing machine \cite{turing1936computable,Currin_2017_interface}. The solution space of NP-complete problems grows super-polynomially with the problem size, which leads to massive computing time even for medium-sized problem and therefore greatly restricts the size of the problem that can be deal with. In contrast to the plight of lacking a practical and efficient computing regime, NP-complete problems are closely related to a wide range of realistic scenarios \cite{Reinelt2003,Smith2005fault,darmann2014subset,kellerer2004multidimensional,Biesner2022solving,eghdami2021ssa}, including transportation, industrial manufacturing, finance, biomedicine and so on, which implies that an acceleration of solving NP-complete problems could lead to a more productive society and might even bring a revolution to future development. 

Over these years, extensive efforts have been dedicated to the exploration of novel computing architectures for NP-complete problems. The emergent approaches which exploit different operational principles or different information carriers have provided more possible ways to cope with the plight, including quantum computation\cite{farhi2001quantum,ChangQuantum2018}, memcomputing\cite{di2013parallel,traversa2015memcomputing,yan2021reconfigurable}, biological computation\cite{Adleman1994molecular,braich2002solution,nicolau2016protein} and optical computing \cite{Wu_2014_Light,vazquez2018hppoe,mcmahon2016fully}. In general, high computing efficiency, high accuracy and programmability are necessary ingredients for a computing architecture to step toward practical application. However,  architectures meeting all the criteria still remains elusive.
Our proof-of-principle demonstration has shown that integrated photonic technology can play a role in building a monolithic computing architecture solving NP-complete problem, which exhibits promising computational potential by taking advantages of the intrinsic properties of photons \cite{xu2020scalable}. Meanwhile, recent progress of integrated photonics enables the realization of programmable optical signal processors \cite{dyakonov2018reconfigurable,bogaerts2020programmable}. The above facts suggest the possibility of constructing a chip-scale NP-complete problem photonic processor fulfilling the practical requirements.

Here, we present a reconfigurable integrated photonic processor for a representative NP-complete problem, the SSP, whose intractability can be utilized to construct attack-resistant cryptosystem \cite{okamoto2000quantum, kate2011generalizing}. The photonic processor is fabricated by femtosecond laser writing techniques\cite{femtowriting}. It consists of on-chip phase shifters and an embedded three-dimensional (3D) waveguide network made of 1449 standardized modules. We map the SSP to the waveguide network, and the incident photons travel in the network to perform parallel computation. The optional entry and the tunable module of the waveguide network provide multiple degrees of freedom for programming the photonic processor, enabling solving different SSP instances. We have also analyzed the reliability and time-consumption performance of the photonic processor to show the photonic advantages.

\section*{Results} 

\subsection*{Architecture of the reconfigurable photonic processor}

Given a set \textit{S} containing \textit{N} integers, the SSP asks whether there exists a subset of \textit{S} whose sum is equal to target \textit{T}. As presented in Fig. \hyperref[fig1]{1a}, we use a photonic processor to solve the SSP, which is composed of phase shifters deposited on the surface and a buried 3D waveguide network encoding the SSP instance where $S=\{2, 3, 5, 7, 11, 13, 17\}$. Once the coherent light enters the waveguide network, the photonic processor is activated to start a computation. Photons contained in the light beam propagate under the regulation of the waveguide network, exploring all the possible paths towards the output ports in a parallel manner. The arrival or absence of photons at the output are read out by one-shot imaging, giving a YES or NO answer to the SSP, respectively.

The detailed architecture of the photonic processor can be understood through an illustration of the constitution of the waveguide network. As shown in  Fig. \hyperref[fig1]{1b}, the waveguide network in Fig. \hyperref[fig1]{1a} can be represented by an abstract network made of lines and nodes. The lines denote optical paths. The five kinds of nodes represent network entry and the functional modules, i.e., fixed split junction, variable split junction, pass junctions and converge junctions. Photons are launched into the network through one of the pink diamond nodes (network entries). At black hexagonal nodes (fixed split junctions, see Fig. \hyperref[fig1]{1c} for physical structure), photons are equally divided into two portions, which then proceed in vertical (i.e., \textit{x}) and diagonal directions, respectively. In the case of yellow hexagonal nodes (variable split junctions, see Fig. \hyperref[fig1]{1d} for physical structure), photons can be split with any specified ratio $\eta:1-\eta$ $(0\leq \eta \leq 1)$ by properly setting the phase shifters (see Methods). Similarly, the split light propagates vertically and diagonally. Blue circular nodes (pass junctions, see Fig. \hyperref[fig1]{1e} for physical structure) enable photons to move forward along the original direction, which is realized by 3D crossing structures difficult to fabricate with traditional lithography. At the end of the network, brown square nodes (converge junctions, see Fig. \hyperref[fig1]{1f} for physical structure) gather together photons from different paths.

The network encodes the SSP according to the following rules. First, hexagonal-node block and circular-node block alternate appear for  $N$ (the number of elements of $S$) times. Second, the vertical distance between two adjacent rows of hexagonal nodes is equal to the element in the set $S$, as denoted by the integers on the left. The distance is measured as the number of nodes. Note that the hexagonal node in a latter row is counted in while the one in a former row is not. Second, the diagonal movement of photons means including an element into the summation while the vertical movement means the opposite. Last, the position of the output signals represents the ultimate sums, which are denoted by the output port number. For example, the path highlighted in pink indicates that elements 3, 5, 11 and 13 are included into the summation, whose value is 32. Besides the mathematic mapping, the network is physically implemented by femtosecond laser writing techniques (see Methods). Meanwhile, the physical structures of the functional modules are elaborately designed and optimized (see Supplementary Section I-II).  

The foundation of reconfiguring the photonic processor is the optional network entry and the tunable functional module, variable split junction. In general case, a photonic processor is initially designed for an SSP instance where $S=\{X_{1}, X_{2}, ..., X_{N}\}$. As illustrated in Fig. \hyperref[fig1]{1g}, there are different paths to achieve reconfiguration of the photonic processor. First, by switching to a different entry, like Entry $i$, we can program the photonic processor to solve another SSP instance where $S=\{X_{i}, X_{i+1}, ..., X_{N}\}$. The reason is that the local network encoding the first $i-1$ elements is bypassed, preventing these elements from participating in the computation. Second, we can choose to delete or keep the element $X_{j}$ by properly setting the working modes of the $j_{th}$ row of variable split junctions, which can be understood through the following deduction. As introduced above, the reflectivity $\eta$ of variable split junctions is tunable. Therefore, we can set the $j_{th}$ row of variable split junctions to total transmission mode ($\eta=0$) or total reflection mode ($\eta=1$), depending on their specific location, to completely transfer the arriving photons to vertical paths. On this occasion, there is a zero probability of including the element $X_{j}$ into any summation. Namely, $X_{j}$ is removed out of the computation. On the contrary, $X_{j}$ is retained when the variable split junctions work in balance mode ($\eta=0.5$). In summary, variable split junctions can be used to decide whether to remove an element, therefore allowing to program the photonic processor. Finally, the two aforementioned methods can be also applied at the same time. For a fully reconfigurable photonic processor (i.e., every split junction is variable), it allows, in principle, $2^{N}$ different configurations at most, implying the potential versatility of the proposed computing architecture. 

\subsection*{Reconfigurability and reliability}
We experimentally investigate the reconfigurability and reliability of the implemented photonic processor, in which the second row of split junctions are variable as depicted in Fig. \hyperref[fig1]{1b} (see Supplementary Section III for the experimental setup). To correctly set the working modes of the variable split junctions, we first characterize their optical response to the dissipated power of the phase shifters (see Methods). The response curves are well consistent with the theoretical expectation, allowing us to easily identify the three working modes (see Supplementary Section IV). 

We achieve programming the photonic processor to solve the SSP instance where $S=\{2, 3, 5, 7, 11, 13, 17\}$ by choosing Entry 1 and setting the variable split junctions to balance mode. With a 810 nm laser coupled into the photonic processor, the computation is started. The evolution results of the light appear as a line of spots (Fig. \hyperref[fig2]{2a}), which certify the existence of the corresponding subset sums (i.e., the numbers below the spots). In other words, the appearance of a spot represents that there exists a subset of $S$ whose sum is equal to the number, and the missing of a spot denotes the opposite case. Compared with the benchmark results attained by exhaustive enumeration, all the observed spots are valid certifications and meanwhile they completely reveal all the possible subset sums, suggesting excellent accuracy of the photonic computing.  

The experimental evolution results are further investigated through an analysis of the intensity distribution, as shown in Fig. \hyperref[fig2]{2b}. The theoretical intensity distribution is obtained based on an ideal model and thus can be treated as a benchmark result (see Methods). In theoretical regime, any signal of nonzero intensity denotes the existence of a subset sum. Whereas, it is not the case in the experiment due to inevitable environmental noise and fabrication imperfection. Nevertheless, we can correctly classify the experimental signals into valid and invalid certifications by applying a reasonable intensity threshold. If the signal has an intensity beyond the threshold, it is identified as a valid certification. Otherwise, it is invalid (highlighted by white solidus pattern).  As indicated by the band filled with gray solidus, the tolerance interval for the threshold is relatively large (with an upper bound of 0.00143 and a lower bound of 0.00027), further confirming the reliability of our photonic processor. 

We are also able to program the photonic processor for a different SSP instance where $S=\{2, 5, 7, 11, 13, 17\}$ by tuning the working modes of the variable split junctions (see Methods). Entry 1 still serves as the input port. Similar to the previous case, the computation outcomes are of high accuracy, as demonstrated by the experimental evolution results (Fig. \hyperref[fig2]{2c}) and the intensity distribution (Fig. \hyperref[fig2]{2d}). In addition, the photonic processor is capable of dealing with more SSP instances by using other network entries for photon injection (see Methods). Figs. \hyperref[fig3]{3a}, \hyperref[fig3]{3c} and \hyperref[fig3]{3d} present the experimental evolution results when the light is injected through Entry 2, Entry 3 and Entry 4, respectively. More results can be found in Supplementary Section V. It should be noticed that, in all the cases, the experimental evolution results are in accordance with the benchmark results acquired by exhaustive enumeration. Furthermore, the tolerance intervals of the thresholds applicable in our experiments are considerably large (Fig. \hyperref[fig3]{3b} and Supplementary Figs. S6-S7), owing to the good experimental signal-to-noise ratio. These facts indicate the achievement of solving multiple SSP instances on the photonic proccesor with high accuracy, verifying the reconfigurability and strong reliability of the photonic computing architecture.

\subsection*{Time-consumption advantage}
Computing time is one of the most critical performances of a computing architecture. We investigate the time consumption of our photonic processor by comparing with representative electronic processors. We define the computing time of the photonic processor as the propagation time of photons in the longest path of the waveguide network. It is obtained through dividing the length of the longest path by the propagation speed of light in the waveguide (see Methods) \cite{xu2020scalable}. Owing to the parallel working manner, the photonic processor is able to give all the possible subset sums at a time, which, to some extent, is equivalent to simultaneously solve a series of SSP instances whose target $T$ is different. For a fair comparison, electronic processors are considered to search the entire solution space to solve the SSP, accompanied with the acquirement of all the subset sums. The computing time of electronic processors is estimated by dividing the total number of arithmetic operations by the floating point operations per second (FLOPS) \cite{flops}.

Fig. \hyperref[fig4]{4a} displays the estimated computing time in the case of successive primes. We find that, at the very beginning, the photonic processor is comparable to Intel Pentium III released in 2001 \cite{IntelPentiumProcessor}, whereas lags behind  Intel Core i7-11370H and i7-1160G7 \cite{IntelCoreProcessor}, the electronic processors launched in recent years. However, as the problem size increases, the photonic processor shows a trend of outperforming all the rivals. We magnify the curves encircled by dashed lines in Fig. \hyperref[fig4]{4a} and find that in our experimental demonstration of instance $S=\{2, 3, 5, 7, 11, 13, 17\}$, the photonic processor has already surpassed all the electronic rivals, as shown in Fig. \hyperref[fig4]{4b}. Specifically, the photonic processor is several orders of magnitude faster than Intel Pentium III and several times faster than the other two rivals. Moreover, the photonic superiority is reinforced with the growth of problem size, showing considerable competitiveness. It should be noticed that the time-consumption advantage of the photonic processor is achieved with classical light, which implies another possible way towards computational advantage in addition to quantum speed-up. In fact, an injection of quantum light into our photonic processor cannot bring computational acceleration despite the demonstrated quantum advantage  \cite{zhong2020quantum, GAO2022100007, Madsen2022quantum}, the reason for which is that the latency arising from photon emission (i.e., quantum light emits only a few photons at a time) hinders the parallel computation of the photonic processor.

\section*{Discussion}

In summary, we construct a reconfigurable and large-scale photonic processor containing 1449 integrated 3D devices by femtosecond laser writing techniques \cite{xu2021quantum, jiao2021two}. The photonic processor allows solving the SSP, a typical NP-complete problem, and possesses good performances in reconfigurability, reliability and time consumption. Photons with strong robustness act as information carriers. Given the low detectable energy level of photons \cite{hou2021waveguide,liu2022advances}, a coherent light beam could contains enormous amounts of independent information carriers. With the injection of the coherent light, a bunch of photons travel in the photonic processor to search for the solution in parallel.

We successfully program the photonic processor to solve different SSP instances by tunning the working modes of the tunable modules or/and changing the entry. It is worth stressing that, in all the cases, the experimental results agree well with the theory, strongly confirming the reliability of the photonic processor. Furthermore, the photonic processor has been capable of exceeding the recently launched commercial electronic processors in the context of successive primes, suggesting considerable computing potential. The photonic speed-up is attributed to the parallel search of photons, the inherently high propagation speed of light and the confining of light to a limited space via advanced integrated photonic technology.

The reconfigurable photonic computing architecture for the SSP, to the best of our knowledge, is first proposed and experimentally realized. Our experimental investigation verifies the feasibility of the proposal, and the presented core idea can be applied to implementing a fully reconfigurable architecture which in principle allows $2^{N}$ configurations at most. The introduction of reconfigurability lays a solid foundation for building a versatile photonic computing platform, which might play a key role in future supercomputing \cite{caulfield2010future}. First, a large number of different SSP instances can be encoded into a single photonic processor. Second, many SSP-based real-life problems and algorithms \cite{schwerdfeger2016fast,alonistiotis2022approximating}, which usually require programmable hardwares, are possible to be tackled in the framework of the photonic computing architecture possessing a potential of accelerating computation. Last but not least, given the fact that any NP-complete problems can be efficiently reduced to a certain NP-complete problem \cite{garey2002computers,karp1972reducibility}, this photonic processor built for SSP provides a potential platform to deal with a variety of  NP-complete problems, such as exact cover problem \cite{oltean2008exact} and Boolean satisfiability problem \cite{oltean2010optical}.

\section*{Methods}
\noindent \textbf{Tunable splitting ratio of variable split junctions.} The variable split junctions are implemented with 3D Mach-Zender interferometers (MZIs) composed of two modified 50:50 directional couplers and a phase shifter as depicted in Fig. \hyperref[fig1]{1d}, which can be represented by the matrix
$$
U_{MZI}
=
\frac{1}{\sqrt{2}}\left(\begin{array}{cc}
1 & i \\
i & 1
\end{array}\right)\left(\begin{array}{cc}
e^{i \varphi_{t}/2} & 0 \\
0 & e^{-i \varphi_{t}/2}
\end{array}\right) \frac{1}{\sqrt{2}}\left(\begin{array}{ll}
1 & i \\
i & 1
\end{array}\right)
= -i\left(\begin{array}{ll}
\sin \frac{\varphi_{t}}{2} & \cos \frac{\varphi_{t}}{2} \\
\cos \frac{\varphi_{t}}{2} & -\sin \frac{\varphi_{t}}{2}
\end{array}\right),
\eqno{(1)}
$$
where $\varphi_{t}=\varphi+\varphi_{0}$ is the total phase difference between the interferometer arms  ($\varphi$ is induced by the phase shifter, $\varphi_{0}$ is the constant initial phase difference arising from imperfect fabrication). According to eq. (1), the variable split junctions have a reflectivity 
$$\eta=\sin^{2}\frac{\varphi_{t}}{2}=\frac{1-\cos \varphi_{t}}{2}, \eqno{(2)}$$
which is dependent on the phase difference. With an appropriate setting of the applied current of the phase shifter, $\eta$ can vary from 0 to 1.\\

\noindent \textbf{Preparation of high-performance photonic processor.} A realization of the desired photonic processor requires high-quality fabrication of the large-scale 3D waveguide network, well construction of the phase shifters with a tunable range of at least $2\pi$, and an accurate alignment between the waveguide network and the phase shifters. We used the femtosecond laser with a pulse duration of 290 fs, a repetition rate of 1 MHz and a central wavelength of 513 nm to fabricate the photonic processor. Before focused by a 50$\times$ objective, the laser was locked by a beam-pointing stabilizer to further enhance fabrication precision. The sample (Corning Eagle XG glass) was placed on the 3D translational stage. With the stage moving at 10 mm/s, the laser, shaped by cylindrical lens, radiated into the sample to write the waveguide network at a depth of 55 $\upmu$m to 155 $\upmu$m. The shallow embedment is used to decrease the power consumed by the phase shifters to achieve a $2\pi$ phase shift. A pulse energy of 185 nJ was used. The overlap segment of the waveguides in converge junctions was written twice. During the second writing, a pulse energy of 250 nJ was applied. To avoid a misalignment between the waveguide network and the phase shifters, we inscribed three triangles on the sample surface as reference marks. 

Phase shifters were formed by ablating the thin metal films deposited on the sample surface \cite{Ceccarelli2019}, which was conducted with the same system. A pulse energy of 245 nJ and a translational speed of 5 mm/s were employed. The thin films consist of 2 nm chromium and 100 nm gold, which were successively deposited via electron beam evaporating after the waveguide fabrication. We use the chromium film to enhance the adhesion of the phase shifters, given the fragile bonding between golden film and glass. The phase shifters should contain two pads connecting external power supply, and a resistor heating waveguides. We adopted wide pads ($\sim$ 3 mm $\times$ 2.3 mm) and reasonably narrow resistors ($\sim$ 0.03 mm $\times$ 5 mm) to ensure a good heating efficiency and a consequent large phase-shift range. We carefully aligned the phase shifters with target waveguides with the help of the reference marks, in light of the fact that the misalignment could greatly decrease the heating efficiency and even lead to unwanted crosstalk in the waveguide network with dense layout. \\

\noindent \textbf{Measuring the optical response of variable split junctions.} The optical response of the variable split junctions is characterized as a function of the dissipated power $P$ of the phase shifters. Given eq. (2) and the linear relation between $P$ and $\varphi$ \cite{Ceccarelli2019}, the output intensities of the variable split junctions are supposed to show cosine oscillation with the change of $P$. Since we cannot directly measure the output of the variable split junctions, the characterization is performed on the basis of the whole waveguide network. Generally, a variable split junction is connected to some output ports of the waveguide network while disconnected to the others. For example, according to the network in Fig. 1b, the photons coming from the left variable split junction have a possibility to arrive at port 3 (a connected case) while are impossible to reach port 2 (a disconnected case). We can obtain the optical response of the variable split junctions by monitoring the output intensity at a connected port. Though the magnitude of the intensity at a connected port might be different from that directly measured at the variable split junctions, the relative intensity changes can well reflect the main properties of the optical response of the variable split junctions, enabling us to identify the three working modes. We denote the connected port used for characterization as ``response port".

It should be noticed that  the crosstalk in our experiments is negligible, as discussed in Supplementary Section VI. On this condition, the output intensity at disconnected port, to some extent, reflects the status of the experimental environment and thus can be treated as reference, which plays a role in subsequent data processing. We denote the disconnected port used for characterization as ``reference port". Since the characterization relies on the relative intensity changes, the adoption of a reference port could be beneficial to reducing the error caused by instable experimental environment such as the fluctuating incident light.

During the characterization, we launched a 810 nm laser into the waveguide network through Entry 1, then gradually increased the dissipated power of the phase shifter by changing the applied current, and meanwhile monitored the output of the waveguide network with a charge-coupled device. After that, we divided the output intensity at the response port by the one at the reference port, and plotted the calculated results as a function of the dissipated power (the  results can be found in Supplementary Section IV).\\

\noindent \textbf{Theoretical intensity distribution.} The theoretical intensity distribution is calculated based on an ideal model where energy loss, environmental noise and fabrication imperfection are not considered. Also, all the functional modules operate in perfect conditions. It includes: (i) Fixed split junctions have a reflectivity that is exactly 0.5. (ii) Variable split junctions can be perfectly switched to any of the three working modes. (iii) There is no energy exchange between the waveguides in pass junctions. (iv) In converge junctions, the photons from both vertical and diagonal paths can be fully coupled into the waveguide segment at the end of the junctions. Under the above assumption, the theoretical intensity distribution can be regarded as a benchmark result.\\

\noindent \textbf{Programming the photonic processor.} (i) The configuration for the SSP instance where $S=\{2, 5, 7, 11, 13, 17\}$ is achieved by setting the left and the right variable split junctions to total transmission mode and total reflection mode, respectively. The working mode of the middle variable split junction makes no difference to the computing results. Meanwhile, Entry 1 is used for photon injection. (ii) With a change of the entry, we can program the photonic processor to solve the following SSP instances: $\{3, 5, 7, 11, 13, 17\}$ (Entry 2), $\{5, 7, 11, 13, 17\}$ (Entry 3), $\{7, 11, 13, 17\}$ (Entry 4), $\{11, 13, 17\}$ (Entry 5) and $\{13, 17\}$ (Entry 6). In the case of $\{3, 5, 7, 11, 13, 17\}$, the right variable split junction is switched to balance mode while the working modes of the remaining variable split junctions make no difference to the computing results. In the other cases, the working modes of all the variable split junctions have no influence on the computing results.\\

\noindent \textbf{Estimation of computing time.} Based on the actual geometrical parameters of the waveguide network and the estimated refractive index of laser-written glass \cite{refractiveindex}, we can obtain the computing time of the photonic processor. The photonic processor has a run time of $O(N+q)$ where $q$ is the sum of the largest subset of $S$.\\

\subsection*{Data availability}
The data that support the findings of this study are available from the corresponding authors on reasonable request.

\subsection*{Acknowledgements}
This research is supported by the National Key R\&D Program of China (2019YFA0308700, 2019YFA0
706302 and 2017YFA0303700); National Natural Science Foundation of China (NSFC) (12104299, 61734005, 11761141014,
11690033); Science and Technology Commission of Shanghai Municipality (STCSM)(20JC1416300, 2019SHZDZX01); Shanghai Municipal Education Commission (SMEC)
(2017-01-07-00-02-E00049); China Postdoctoral Science Foundation (2021M692094, 2022T150415). X.-M.J. acknowledges additional support from a Shanghai talent program and support from Zhiyuan Innovative Research Center of Shanghai Jiao Tong University.

\subsection*{Author contributions} 
X.-M.J. supervised the project. X.-Y.X. designed the experiment and fabricated the photonic processor. X.-Y.X., T.-Y.Z. and Z.-W.W. performed the experiment and analyzed the data. X.-Y.X. and X.-M.J. wrote the paper, with input from all the other authors.

\subsection*{Competing interests}
The authors declare no competing interests.

\baselineskip 21pt

\begin{figure*}
\centering
\includegraphics[width=0.95\columnwidth]{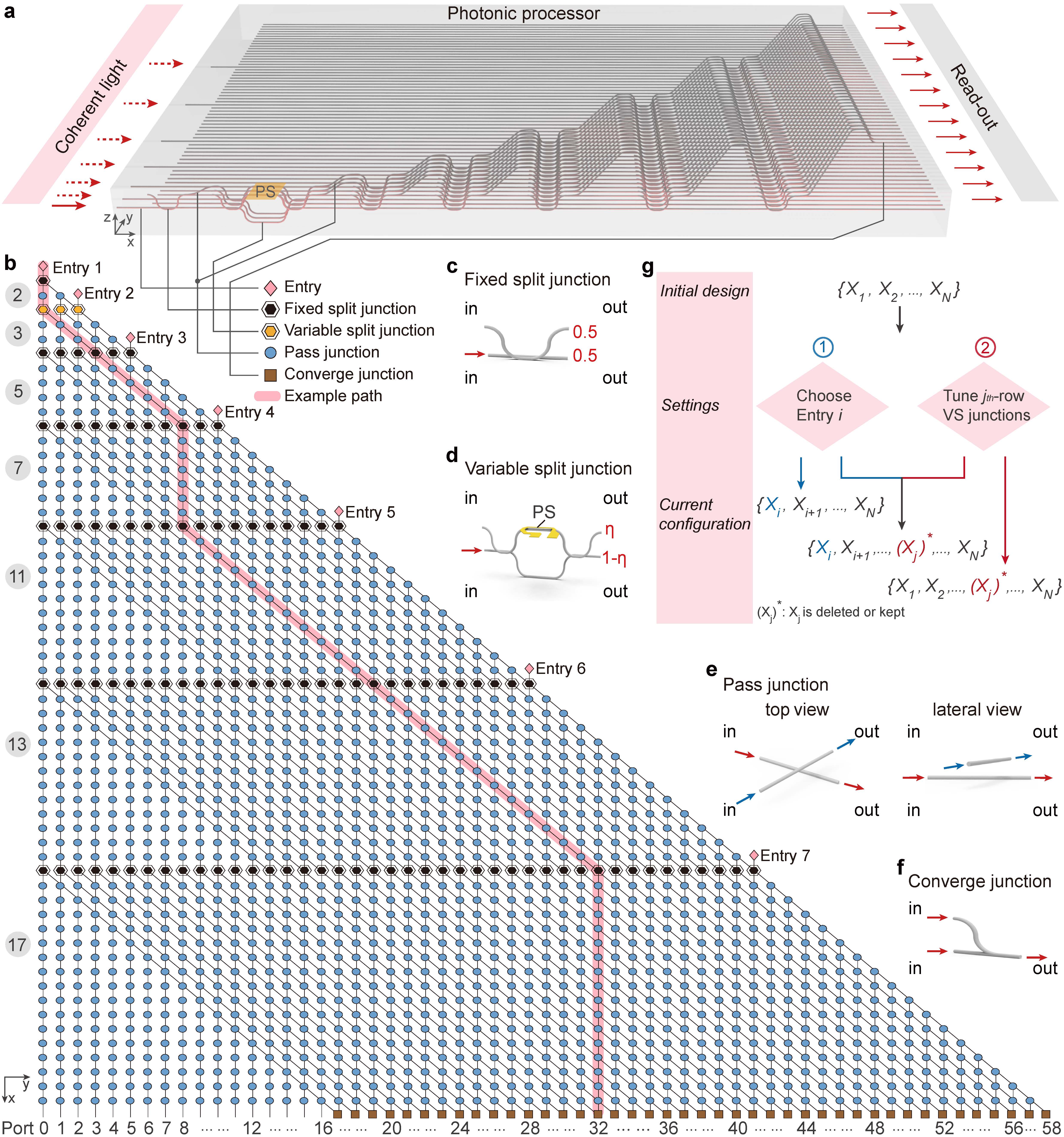}
\caption{\textbf{Architecture of the photonic processor.} \textbf{a}, The photonic processor consists of phase shifters (PSs) and a waveguide network encoding the SSP instance \{2, 3, 5, 7, 11, 13, 17\}. Coherent light is injected into the network via one of the entries, and  the evolution results are read out to give the solution. \textbf{b}, Waveguide network in \textbf{a} can be represented by a network where lines denote optical paths, and nodes denote entry and four kinds of functional modules. The vertical (\textit{x}-direction) distance between two adjacent rows of hexagonal nodes is equal to the elements, as denoted by the integers on the left. Vertical (diagonal) movement of light means excluding (including) an element out of (into) the summation, whose value is denoted by the port number of the output signal. The path highlighted in pink indicates that elements 3, 5, 11 and 13 are included, resulting in a sum 32. \textbf{c}, Fixed split junctions equally split the light. \textbf{d}, Variable split (VS)  junctions can split the light with any specified ratio $\eta:1-\eta$ $(0\leq \eta \leq 1)$ by properly setting the phase shifters. \textbf{e}, Pass junctions preserve the original propagation of light. \textbf{f}, Converge junctions gather together light from different paths.  \textbf{g}, A photonic processor initially designed for $\{X_{1}, X_{2},..., X_{N}\}$ can be programmed by changing entry or/and tuning variable split junctions.}
\label{fig1}
\end{figure*}
%

\begin{figure*}
\centering
\includegraphics[width=1\columnwidth]{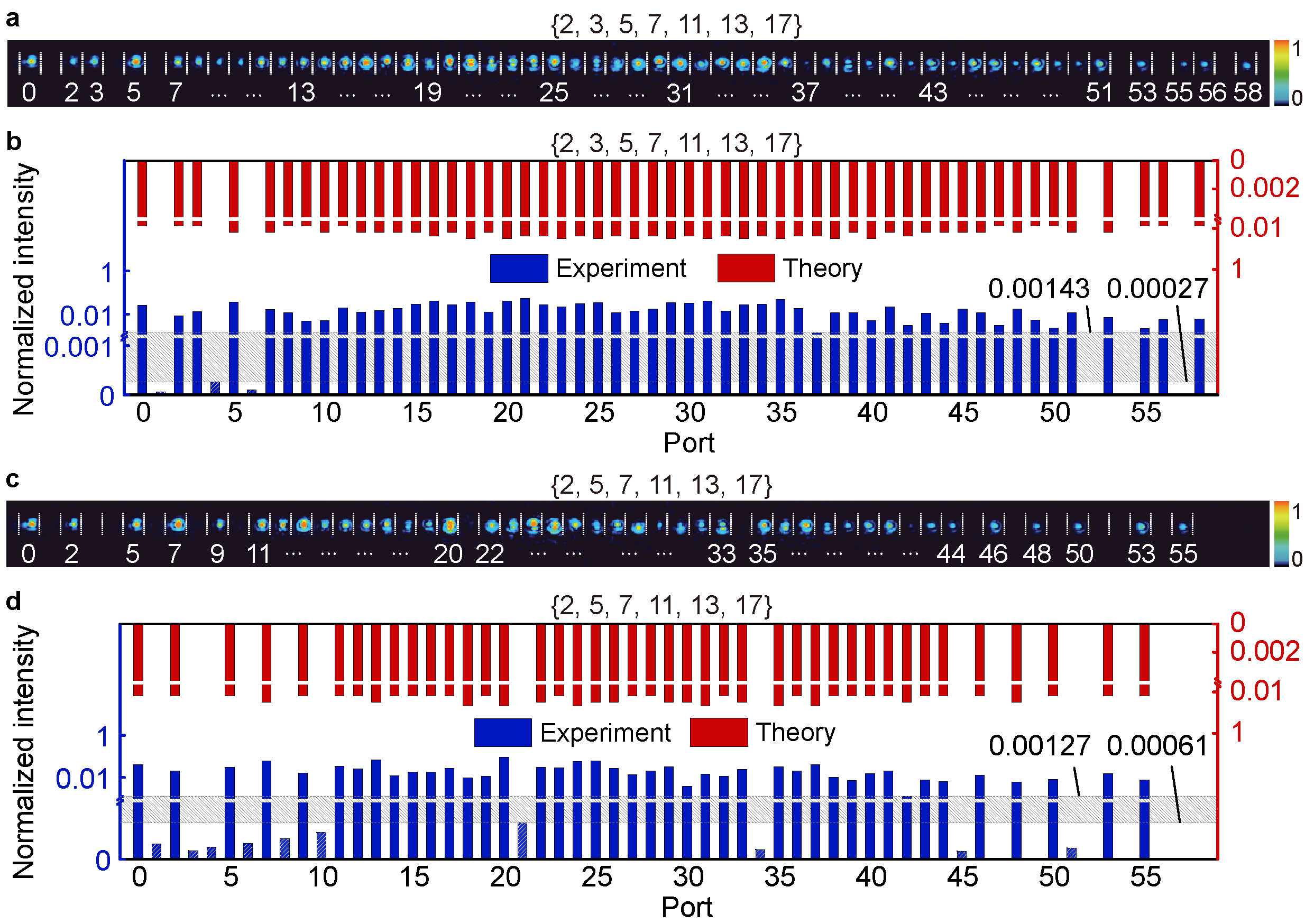}
\caption{\textbf{Computing results of the cases \{2, 3, 5, 7, 11, 13, 17\} and \{2, 5, 7, 11, 13, 17\}.} \textbf{a} and \textbf{c}, The experimental read-out displays as a line of spots, which certify the existence of the corresponding subset sums (i.e., the numbers below the spots). \textbf{b} and \textbf{d}, The experimental and theoretical intensity distribution. Axis break is used for the joint display of logarithmic coordinates and zero intensity. In the theoretical cases, nonzero intensity certify the existence of a subset sum. By applying a reasonable intensity threshold, the experimental signals can be correctly classified into valid (beyond the threshold) and invalid certifications (below the threshold, highlighted by white solidus pattern). The tolerance intervals of the thresholds are marked by the bands filled with gray solidus, revealing the upper bounds and the lower bounds.
}
\label{fig2}
\end{figure*}

\begin{figure*}
\centering
\includegraphics[width=1 \columnwidth]{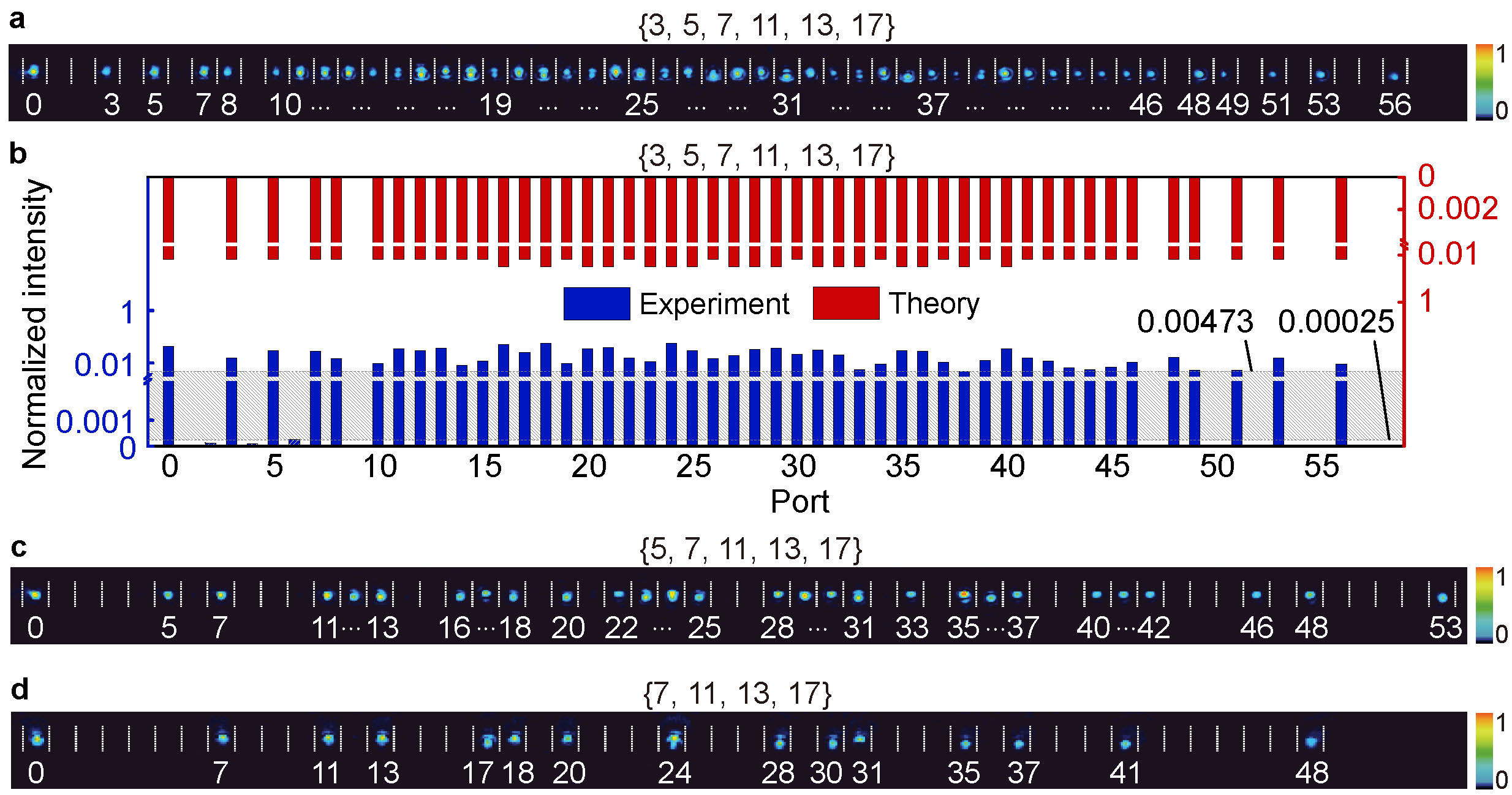}
\caption{\textbf{Computing results of the cases \{3, 5, 7, 11, 13, 17\}, \{5, 7, 11, 13, 17\} and \{7, 11, 13, 17\}.} \textbf{a}, The experimental read-out of the case \{3, 5, 7, 11, 13, 17\} and \textbf{b}, the corresponding intensity distribution. The threshold applicable in our experiments has a  considerably large tolerance interval, whose upper bound and lower bound is 0.00473 and 0.00025, respectively, as indicated by the band filled with solidus. \textbf{c-d} The experimental read-outs of the cases \{5, 7, 11, 13, 17\} and \{7, 11, 13, 17\}. The corresponding intensity distribution is presented in Supplementary Fig. S6.}
\label{fig3}
\end{figure*}

\begin{figure*}
\centering
\includegraphics[width=1 \columnwidth]{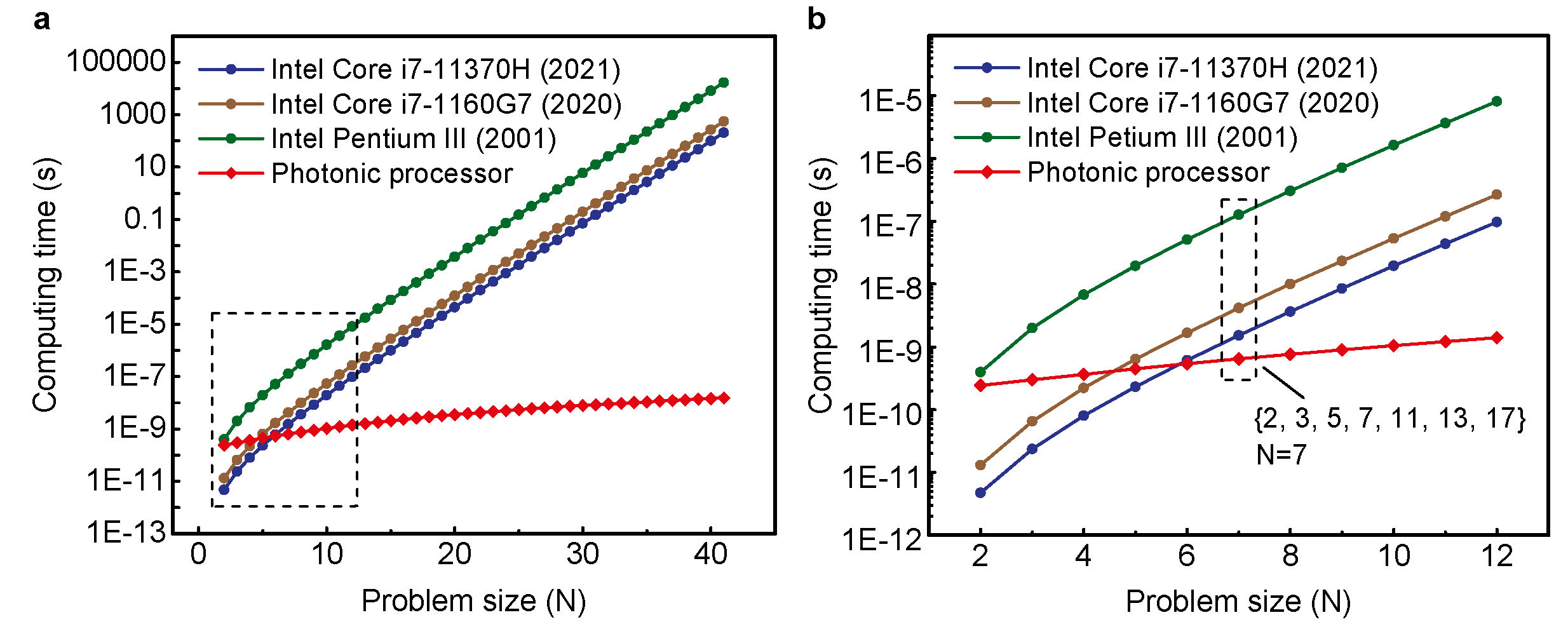}
\caption{\textbf{Time-consumption performance.} \textbf{a}, In the case of successive primes \{2, 3, 5, 7, ...\}, the computing time of our photonic processor is estimated and compared with the representative electronic processors, which are released in 2001, 2020 and 2021, respectively. The electronic processors, which search the entire solution space to solve the SSP, are superior to the photonic processor only at the early stage. With the increase of problem size, the photonic processor has an increasingly obvious advantage over the electronic rivals. A magnification of the curves encircled by dashed lines is exhibited in \textbf{b}. Clearly, our photonic processor has already outperformed all the electronic processors in the experimental demonstration of the SSP instance where S=\{2, 3, 5, 7, 11, 13, 17\}. }
\label{fig4}
\end{figure*}

\clearpage 


\renewcommand{\thesubsection}{S.\Roman{subsection}} 
\renewcommand{\thefigure}{S\arabic{figure}}
\renewcommand{\thetable}{S\Roman{table}}
\setcounter{figure}{0}
\renewcommand{\theequation}{S\arabic{equation}}

\section*{Supplementary Information: A Reconfigurable Photonic Processor for NP-complete Problems}

\baselineskip 24pt  

\subsection{Structure designs of the functional modules}
\begin{figure*}[htbp]
\centering
\includegraphics[width=0.85\columnwidth]{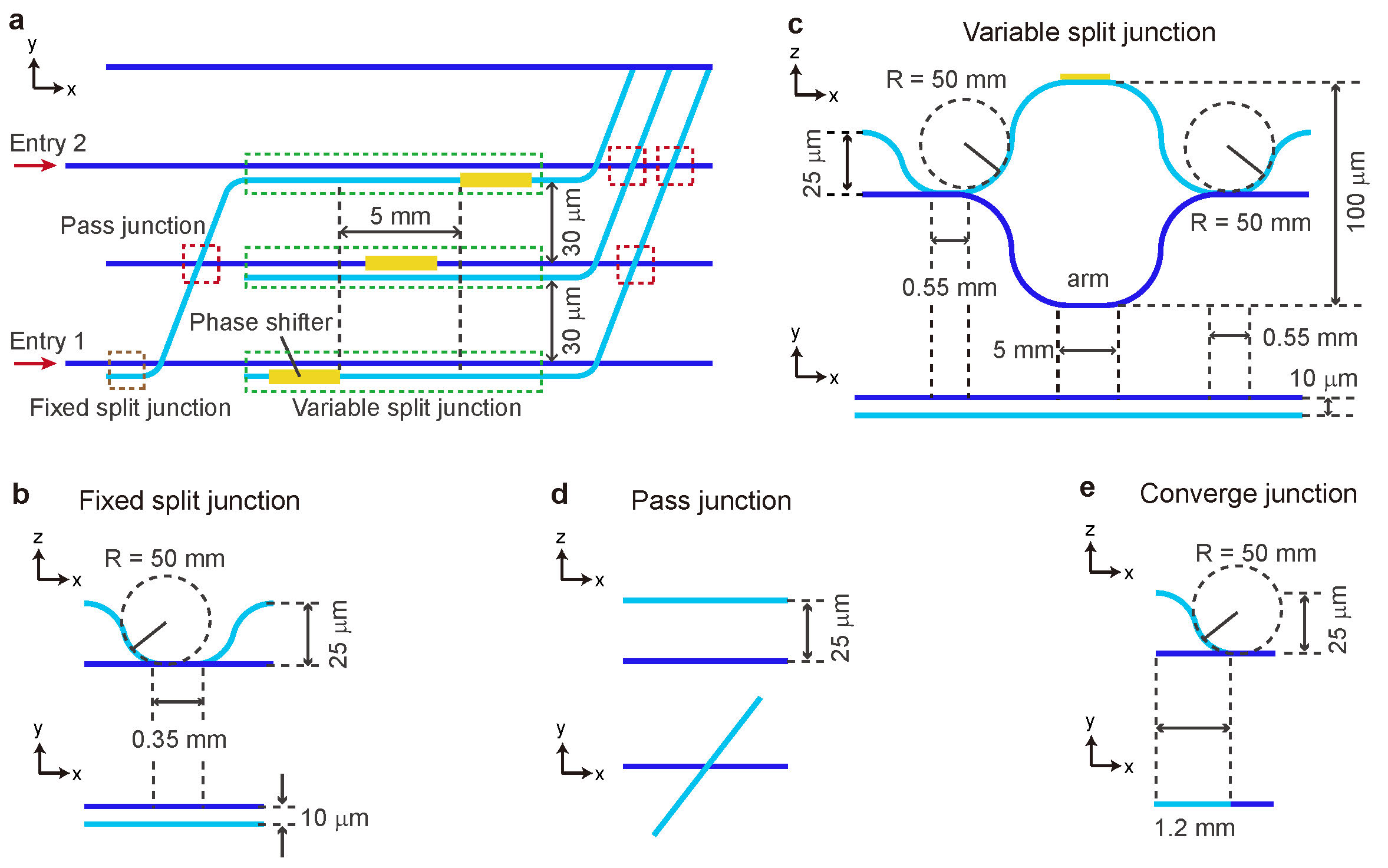}
\caption{\textbf{An overview of the top corner of the waveguide network, and structure designs of the functional modules.} \textbf{a}, The \textit{x}-\textit{y} view of the top corner of the waveguide network, which constitutes standardized functional modules, fixed split junctions, variable split junctions and pass junctions. \textbf{b}-\textbf{e}, The \textit{x}-\textit{z} and \textit{x}-\textit{y} views of fixed split junctions, variable split junctions, pass junctions and converge junctions.}
\label{figs1}
\end{figure*}

We first give an overview of the top corner of the waveguide network to show the specific layout of the functional modules (Fig. \hyperref[figs1]{S1a}), and then illustrate the structure designs of the functional modules in detail (Figs. \hyperref[figs1]{S1b}-\hyperref[figs1]{S1e}). As presented in Fig. \hyperref[figs1]{S1a}, the spacing between a blue waveguide and its next nearest cyan waveguide is 30 $\upmu$m, at a distance of which the energy exchange through evanescent coupling is negligible (see Supplementary Section II). The phase shifters in  the variable split junctions are spaced in the \textit{x} direction. The separation  between the phase shifters deposited above the lower and the upper variable split junctions is 5 mm, which is large enough to eliminate the thermal crosstalk between them, as presented in the Supplementary Section VI. Light is coupled into the network through the entries, located at the front end of the photonic processor.

Fixed split junctions (encircled by brown dashed lines in Fig. \hyperref[figs1]{S1a}) are actually modified three-dimensional directional couplers which have a bending radius of 50 mm, as displayed in Fig. \hyperref[figs1]{S1b}. A coupling length of 0.35 mm and a coupling distance of 10 $\upmu$m are utilized to achieve a balanced splitting ratio (see Supplementary Section II). Also, the \textit{z}-direction spacing between the two input ports (or between the two output ports) is deliberately set to 25 $\upmu$m to realize decoupling (see Supplementary Section II). 

Variable split junctions (encircled by green dashed lines in Fig. \hyperref[figs1]{S1a}) are realized by Mach-Zehnder interferometers made of two cascaded three-dimensional 50:50 directional couplers (Fig. \hyperref[figs1]{S1c}). The directional couplers are the mirror image of each other and they are connected by the interferometer arms with a length of 5 mm. A bending radius of 50 mm, a coupling length of 0.55 mm and a coupling distance of 10 $\upmu$m are used to achieve a balanced splitting ratio. Compared with fixed split junctions, the \textit{z}-direction spacing between the two outputs of the first coupler (or between the two inputs of the second coupler) is increased to 100 $\upmu$m, which is aimed at enhancing the temperature difference between the interferometer arms. 

Pass junctions (encircled by red dashed lines in Fig. \hyperref[figs1]{S1a}) is composed of two completely separate waveguides.  As exhibited in Fig. \hyperref[figs1]{S1d}, the waveguides look like intersecting with each other in the \textit{x}-\textit{y} view, but they are actually written at different depths of the photonic processor (as displayed in the \textit{x}-\textit{z} view). The distance between them in the \textit{z} direction is 25 $\upmu$m, which allows to decoupling the two waveguides. 

Converge junctions are located at the back end of the whole waveguide network, which are not depicted in Fig. \hyperref[figs1]{S1a}. The two waveguides in converge junctions are separated at the beginning but finally merge into one whole (Fig. \hyperref[figs1]{S1e}), which is designed to gather together photons from two different paths.

\newpage

\subsection{Optimization of the photonic processor}
The photonic processor is optimized by elaborately selecting the decoupling distance between neighboring waveguides and the coupling length of the functional modules. The decoupling distance is chosen based on the model that consists of two parallel straight waveguides. Different groups of waveguide, with increasing waveguide spacing, are fabricated. A 810 nm laser is coupled into one of waveguides. For decoupling waveguides, there is no energy exchange between them. Otherwise, there is a transmission of the light from the input waveguide to the other one.

\begin{figure*}[htbp]
\centering
\includegraphics[width=0.65 \columnwidth]{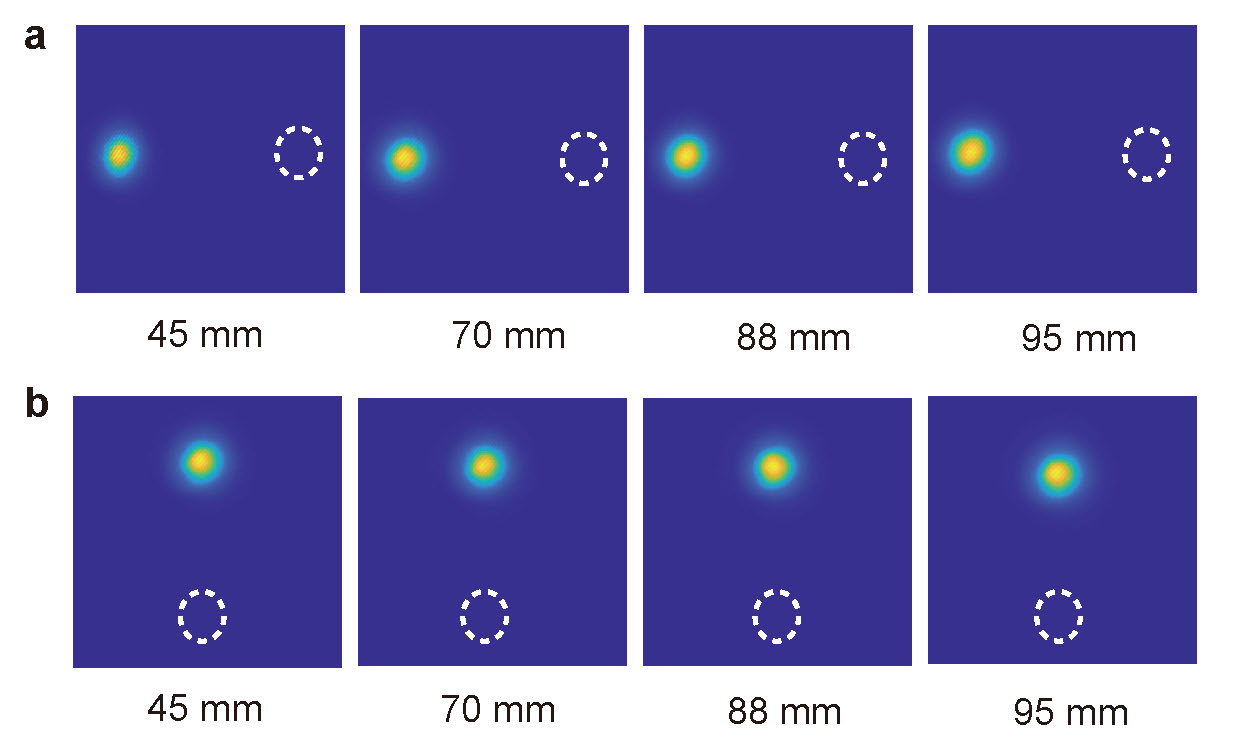}
\caption{\textbf{The intensity distribution of two decoupling waveguides.} \textbf{a}, The waveguides are separated in the \textit{y} direction at a constant distance of 30 $\upmu$m, whereas the waveguide length varies from 45 mm to 95 mm.  A 810 nm laser is coupled into the left waveguide. The position of the right waveguide is marked by dashed lines. \textbf{b}, The waveguides are separated in the \textit{z} direction at a constant distance of 25 $\upmu$m and the waveguide length varies from 45 mm to 95 mm. A 810 nm laser is coupled into the upper waveguide. The position of the lower waveguide is marked by dashed lines.}
\label{figs2}
\end{figure*}

Fig. \hyperref[figs2]{S2a} presents the experimental results when the \textit{y}-direction waveguide spacing is 30 $\upmu$m. We can clearly see that the output signal always stays in the input waveguide (the other waveguide is marked by dashed lines), though the waveguide length varies at a very large range (from 45 mm to 95 mm). Furthermore, the intensity distribution remains unchanged despite the increase of waveguide length. The intensity ratio between the two outputs is up to several hundred. The phenomenon reveals that the evanescent coupling between the waveguides is negligible. Therefore, 30 $\upmu$m is used as the \textit{y}-direction decoupling distance. Similar results are obtained when the \textit{z}-direction waveguide spacing is 25 $\upmu$m (Fig. \hyperref[figs2]{S2b}). As a result,  25 $\upmu$m is used as the \textit{z}-direction decoupling distance. The difference between the decoupling distance in the \textit{y} and \textit{z} directions is attributed to the direction-dependent coupling coefficient of the waveguides.

\begin{figure*}[htp]
\centering
\includegraphics[width=0.65 \columnwidth]{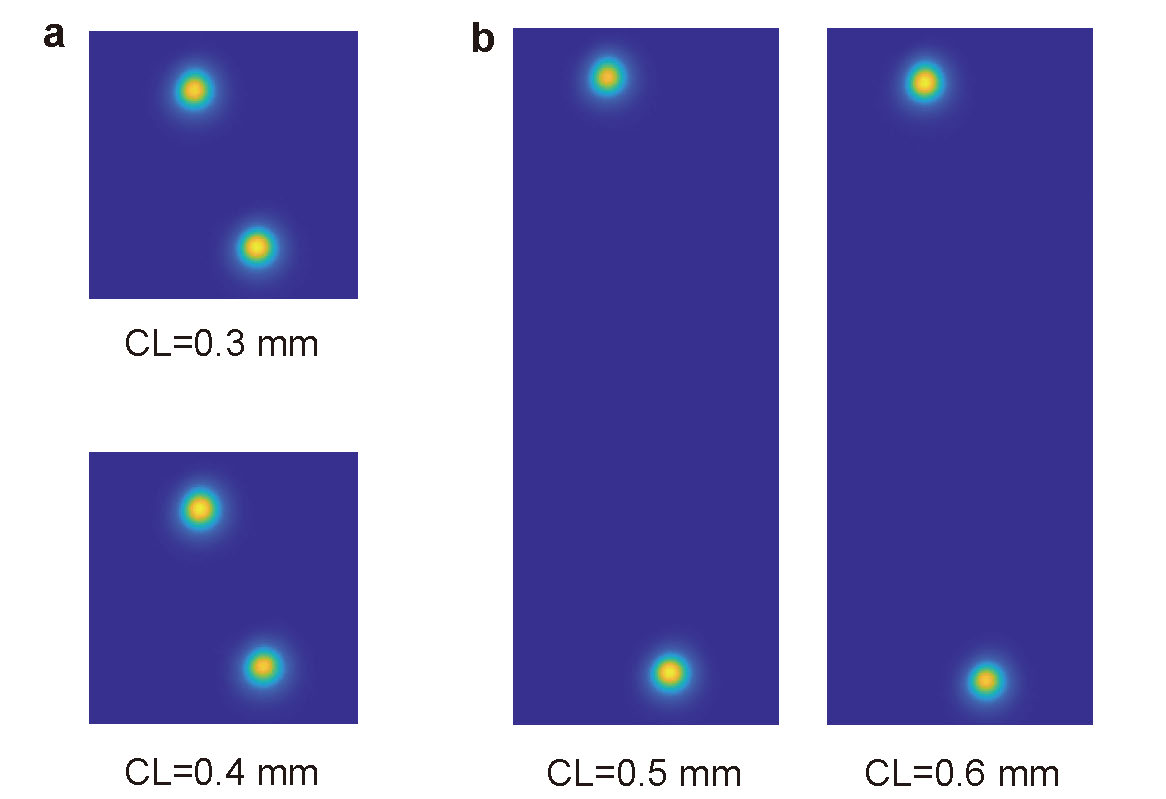}
\caption{\textbf{Intensity distribution of fixed and variable split junctions.} \textbf{a}, The intensity distribution of fixed split junctions when the coupling length (CL) is 0.3 mm and 0.4 mm, respectively. \textbf{b}, The intensity distribution of the directional coupler in variable split junctions when the coupling length (CL) is 0.5 mm and 0.6 mm, respectively.}
\label{figs3}
\end{figure*}

Based on the above decoupling distance, we continue to optimize the coupling length of split  junctions. Given that coupling distance and coupling length play a similar role in evanescent coupling, we fix the coupling distance to 10 $\upmu$m while increase the coupling length at a step of 0.1 mm. For fixed split junctions, a coupling length of 0.3 mm (0.4 mm) leads to a splitting ratio slightly higher (lower) than 50:50, as shown in Fig. \hyperref[figs3]{S3a}. Therefore, we apply a coupling length of 0.35 mm to fixed split junctions. For variable split junctions, we are concerned about the splitting ratio of the cascaded directional couplers. As exhibited in Fig. \hyperref[figs3]{S3b}, the coupling lengths 0.5 mm and 0.6 mm, respectively, lead to splitting ratios slightly higher and slightly lower than 50:50. As a consequence, a coupling length of 0.55 mm is applied to variable split junctions.

\clearpage

\subsection{Experimental setup}

\begin{figure*}[htp]
\centering
\includegraphics[width=0.65\columnwidth]{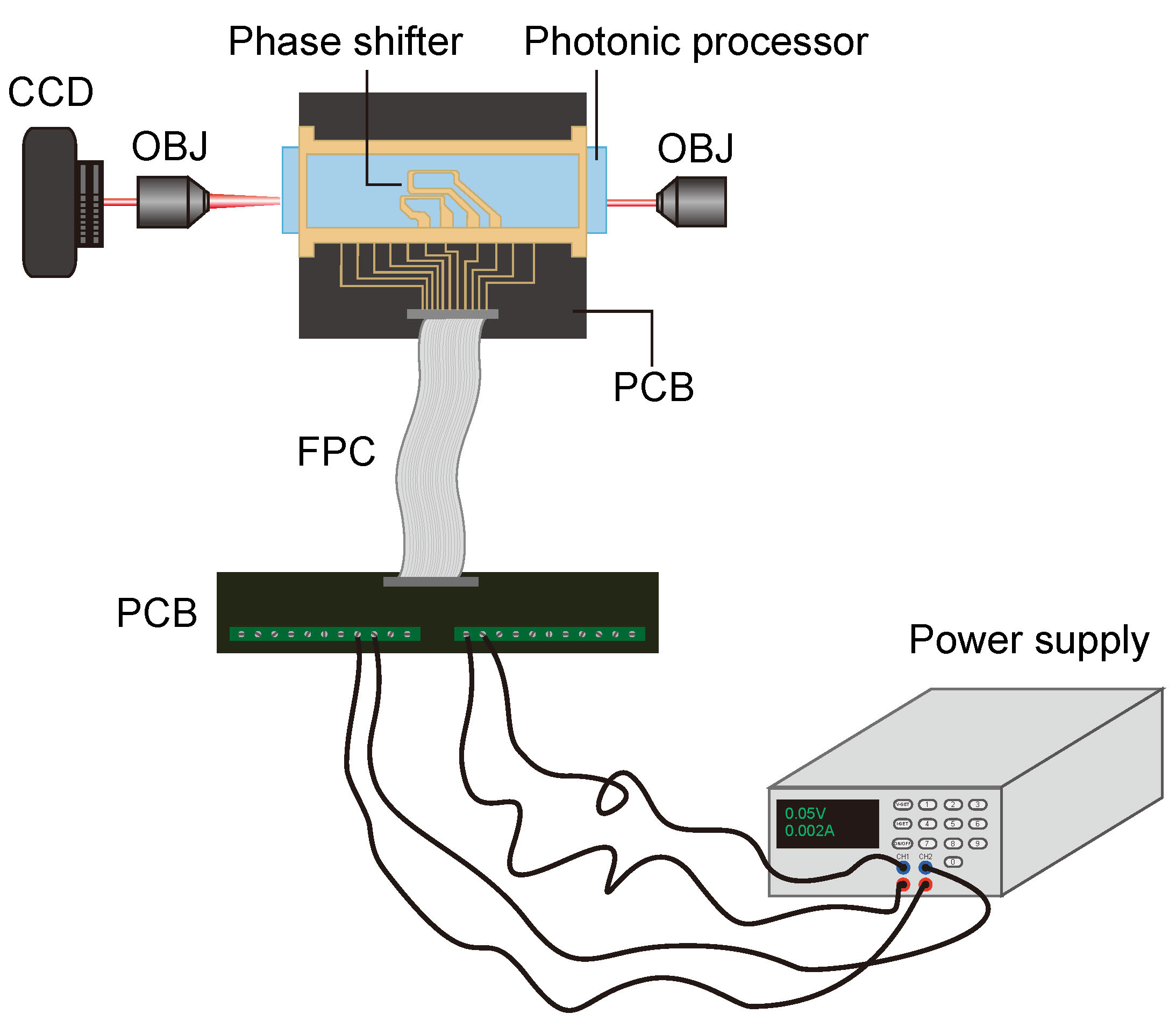}
\caption{\textbf{Experimental setup.} CCD: charge-coupled device; OBJ: objective; PCB: printed circuit board; FPC: flexible printed circuit. }
\label{figs4}
\end{figure*}

As presented in Fig. \hyperref[figs4]{S4}, the photonic processor, with phase shifters deposited on the surface, is bonding to a printed circuit board (PCB). The PCB connects to an external power supply to control the dissipated power of the phase shifters, with the assistance of a flexible printed circuit (FPC) and a second PCB. The power supply is set to constant current mode to eliminate the electrical crosstalk between the phase shifters \cite{Ceccarelli2019}. A laser is coupled into the photonic processor through an objective (OBJ). The experimental results are collected using a charge-coupled device (CCD). It should be noted that the packaged photonic processor and the two objectives are mounted to high-dimensional precision translational stages to ensure accurate optical alignment. Both the characterization of the variable split junctions and the computation of the SSP are carried out with the setup.

\clearpage

\subsection{Optical response of the variable split junctions}
There are three variable split junctions in our photonic processor. According to the network in Fig. \hyperref[fig1]{1b} in the main text, the incident light, in any case, is impossible to go through the variable split junction located in the middle. Therefore, we only characterize the left and the right variable split junctions. In the case of the left one, output ports 2 and 3 serve as ``reference port'' and ``response port'', respectively. The situation for the right one is opposite to the case of the left one. 

As displayed in Fig. \hyperref[figs5.0]{S5}, the optical responses of both the variable split junctions show cosine oscillation, which is consistent with the theoretical expectation. Meanwhile, the three kinds of working modes can be clearly identified. The maximums (minimums) of the response curves correspond to total reflection (transmission) mode where $\eta=1 (\eta=0)$. The median points marked by dashed lines correspond to balance mode where $\eta=0.5$. Note that the phase shifters in the left and the right variable split junctions have a resistance of 71.6 $\Omega$ and 70.5 $\Omega$, respectively.

\begin{figure*}[htp]
\centering
\includegraphics[width=0.85\columnwidth]{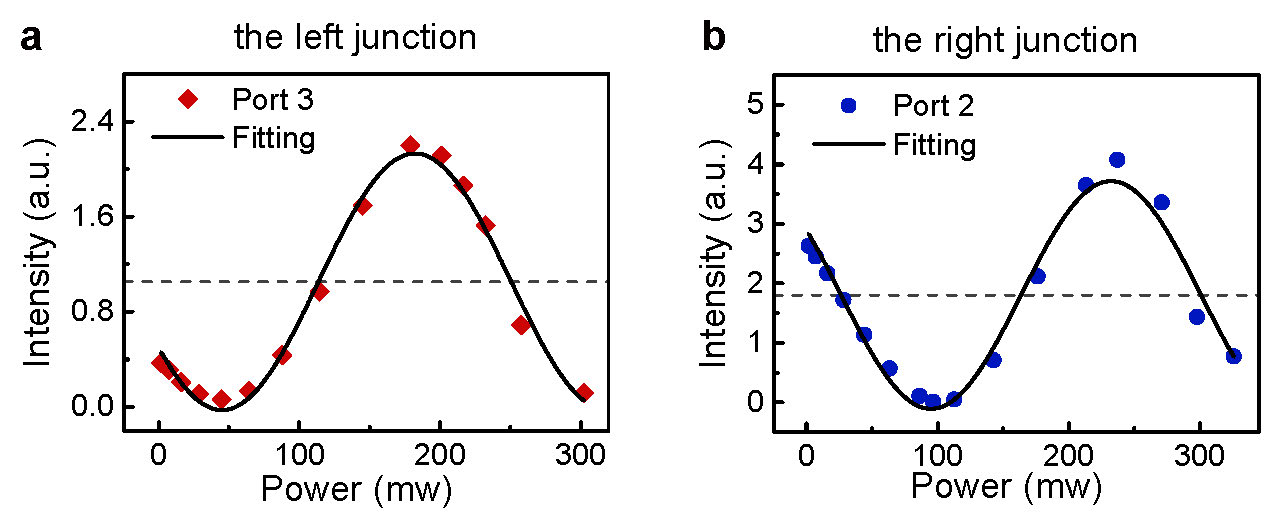}
\caption{\textbf{Measured optical response of the variable split junctions.} The optical response of the left (a) and the right (b) variable split junctions is plotted as a function of the power of the phase shifters. The experimental data are well fitted by a cosine function. The dashed lines mark the median points between the maximums and the minimums. }
\label{figs5}
\end{figure*}

\clearpage

\subsection{Computing results of the photonic processor}
Fig. S6 presents the intensity distribution of the SSP instances where $S=\{5, 7, 11, 13, 17\}$ and  $S=\{7, 11, 13, 17\}$. In both cases, the tolerance intervals of the thresholds have an upper bound that is much larger than the lower bound, as denoted by the bands filled with solidus. Namely, there are a large range of thresholds allowing us to correctly distinguish the valid experimental signals from the invalid ones (highlighted with white solidus pattern). As introduced in the main text, an experimental signal beyond the threshold is identified to be valid certification of the corresponding subset sum. In contrast, an experimental signal below the threshold is considered to be invalid.

\begin{figure*}[htbp]
\centering
\includegraphics[width=1\columnwidth]{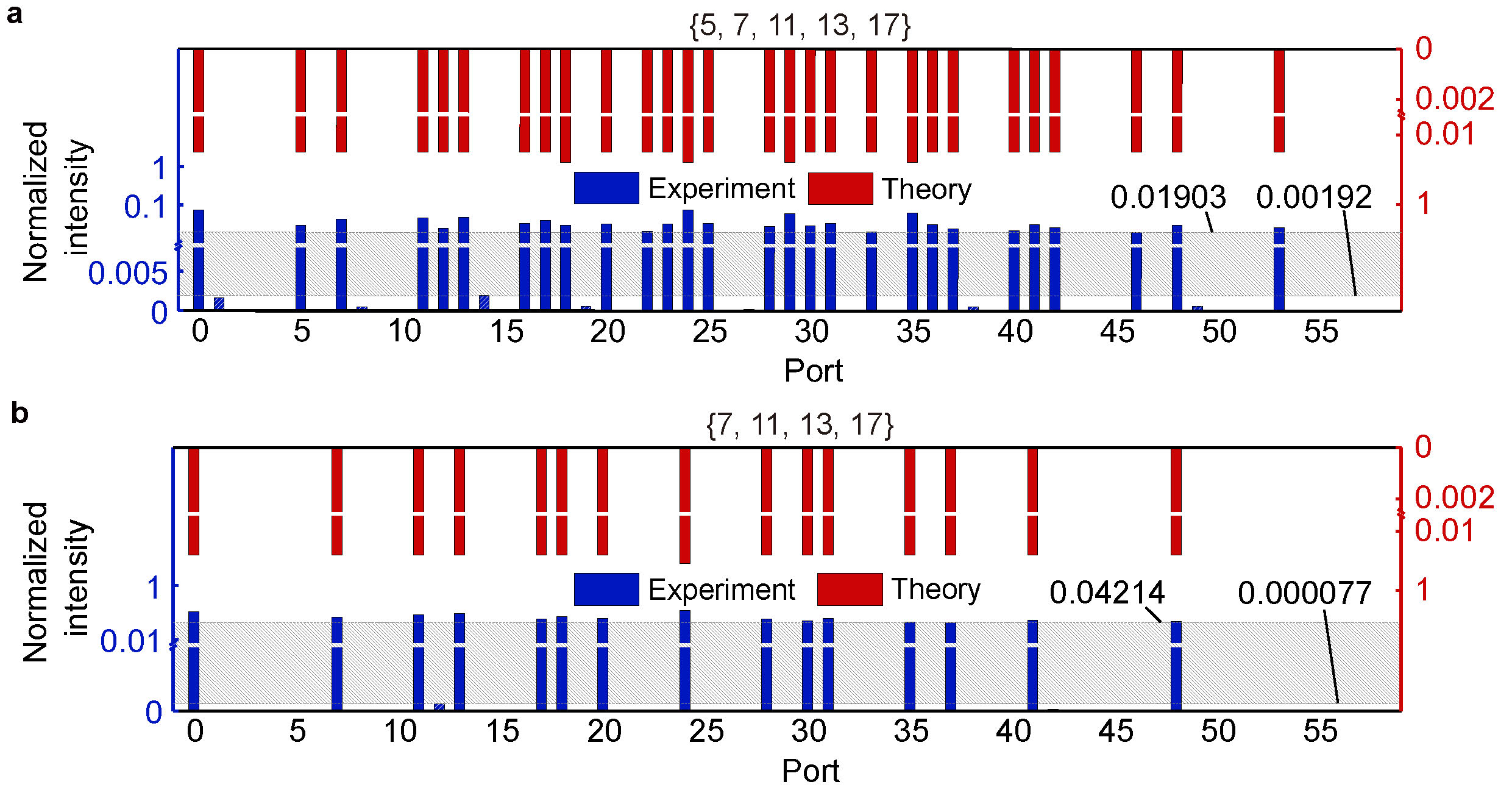}
\caption{\textbf{The intensity distribution of the cases \{5, 7, 11, 13, 17\} and  \{7, 11, 13, 17\}.} \textbf{a}, The tolerance interval of the threshold, marked by the band filled with solidus, has an upper bound of 0.01903 and a lower bound of 0.00192 in the case of \{5, 7, 11, 13, 17\}. \textbf{b}, The tolerance interval of the threshold has an upper bound of 0.04214 and a lower bound of 0.000077 in the case of \{7, 11, 13, 17\}, as the band filled with solidus indicates.}
\label{figs6}
\end{figure*}

More SSP instances can be solved when photons are injected into the photonic processor through a different entry. Figs. S7a and S7c exhibit the experimental evolution results when Entry 5 and Entry 6 act as the input port, respectively. It is found that the computing results agree well with the benchmark results attained by enumeration. Also, the tolerance intervals of the thresholds, shown in Figs. S7b and S7d and marked by the bands filled with solidus, are large enough to accept a lot of thresholds that can correctly separate the valid experimental signals from the invalid ones (highlighted with white solidus pattern).

\begin{figure*}[htbp]
\centering
\includegraphics[width=1\columnwidth]{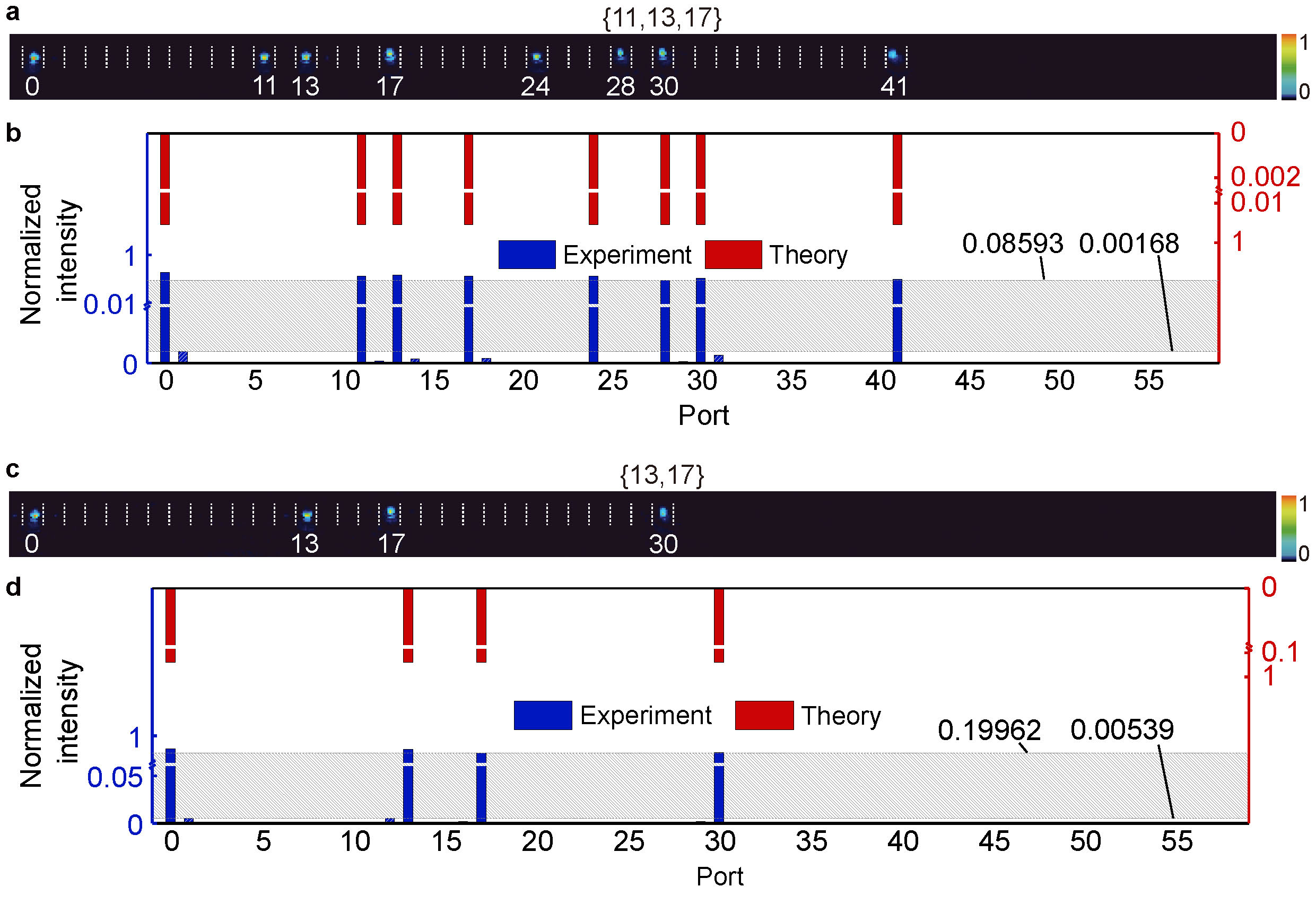}
\caption{\textbf{The computing results of the cases \{11, 13, 17\} and  \{13, 17\}.}  \textbf{a}, The experimental read-out and \textbf{b}, intensity distribution in the case of \{11, 13, 17\}. \textbf{c}, The experimental read-out and \textbf{d}, intensity distribution in the case of \{13, 17\}. The tolerance intervals of the thresholds are indicated by the band filled with solidus.}
\label{figs7}
\end{figure*}

\clearpage
\subsection{Negligible crosstalk}
The phase shifter in the left variable split junction (denoted as VS junction 1) might bring an unwanted phase shift to the right variable split junction (denoted as VS junction 2) and vise versa, since the photonic processor is not a perfect thermal insulator. Based on eq. (2) in the main text and the linear phase-power relation \cite{Ceccarelli2019}, the thermal crosstalk induced by VS junction 1 could result in a change of the output intensity of VS junction 2, which is supposed to be a cosine function of the power of the phase shifter. Therefore, we can evaluate the thermal crosstalk arising from VS junction 1 by measuring the intensity change at the output port that is connected to VS junction 2 while disconnected to VS junction 1. Note that the disconnection means that there does not exist an optical path allowing photons to propagate from the variable split junction to the output port. For example, output port 2 is connected to VS junction 2 but disconnected to VS junction 1, according to Fig. 1b in the main text. 

\begin{figure*}[htp]
\centering
\includegraphics[width=0.45\columnwidth]{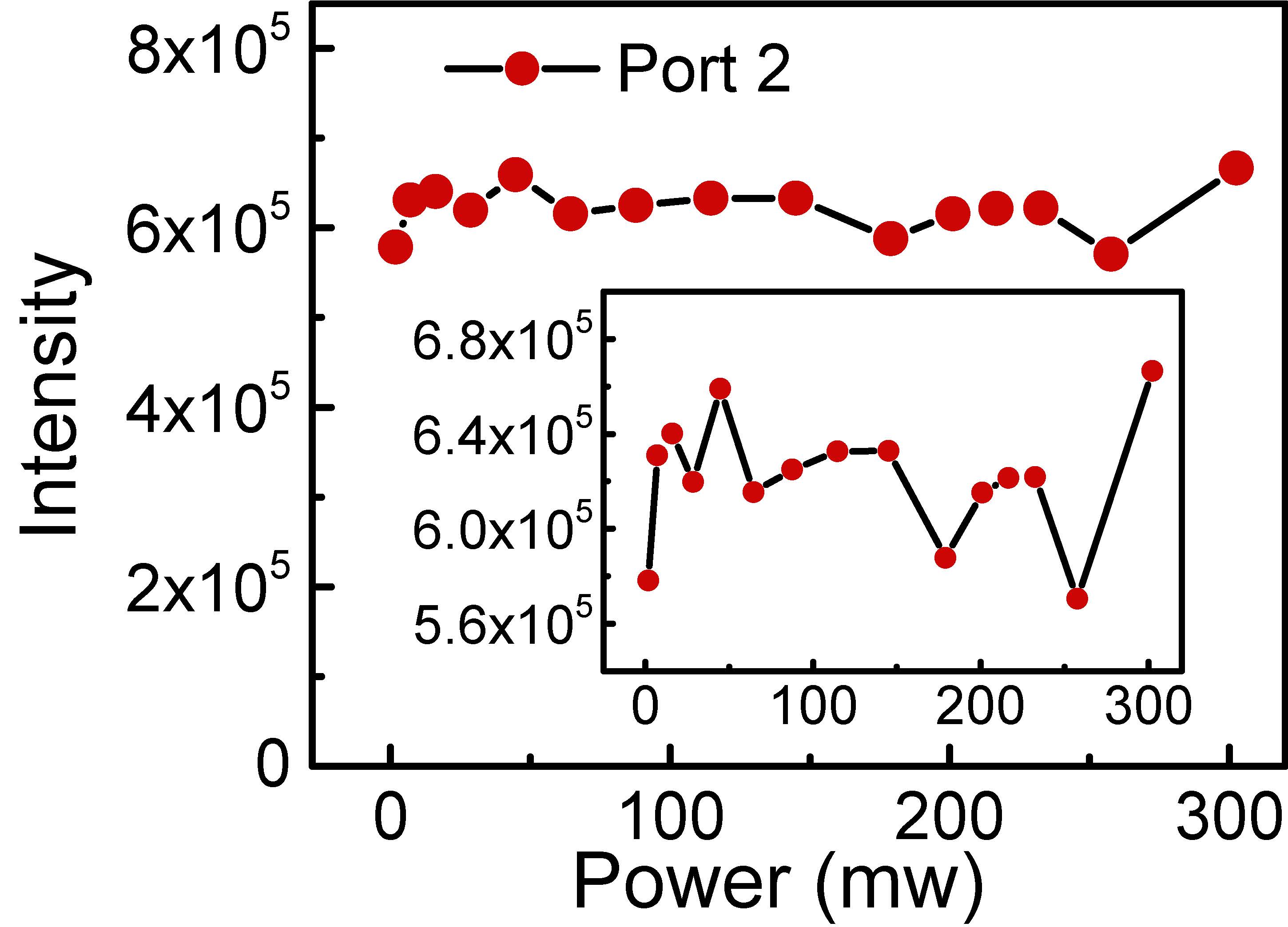}
\caption{\textbf{Intensity at output port 2 during the characterization of VS junction 1.} Output port 2 is disconnected to VS junction 1 while connected to VS junction 2. The intensity is measured when VS junction 1 is characterized and VS junction 2 undergoes a power cut. The irregular and small intensity fluctuation indicates the negligible thermal crosstalk.}
\label{figs8}
\end{figure*}

As demonstrated in Figs. \hyperref[fig2]{S8}, we observe small and irregular intensity fluctuation at output port 2, when VS junction 1 is characterized and VS junction 2 undergoes a power cut. The magnitude of intensity fluctuation is around 14.4\%, which is calculated with the following equation 
$$
\delta=\frac{I_{max}-I_{min}}{I_{max}},
\eqno{(S1)}
$$
where $I_{max}$ and $I_{min}$ are the maximum intensity and the minimum intensity, respectively. In an ideal case, the intensity at output port 2 should be stable. Whereas, in a realistic case, there could be intensity fluctuation, which could be attributed to environmental noise, fabrication imperfectness and unwanted crosstalk caused by the neighboring variable split junction. Obviously, the irregularity of the intensity fluctuation at output port 2 is a strong evidence of the negligibility of the crosstalk induced by VS junction 1. 

We obtain similar results at output port 3 when VS junction 2 is characterized and VS junction 1 undergoes a power cut, as shown in Fig. S9. The magnitude of the intensity fluctuation is about 13.4\%. Note that output port 3 is disconnected to VS junction 2 while connected to VS junction 1. The experimental results verify again that the thermal crosstalk between the variable split  junctions is ignorable.

\begin{figure*}[htp]
\centering
\includegraphics[width=0.45\columnwidth]{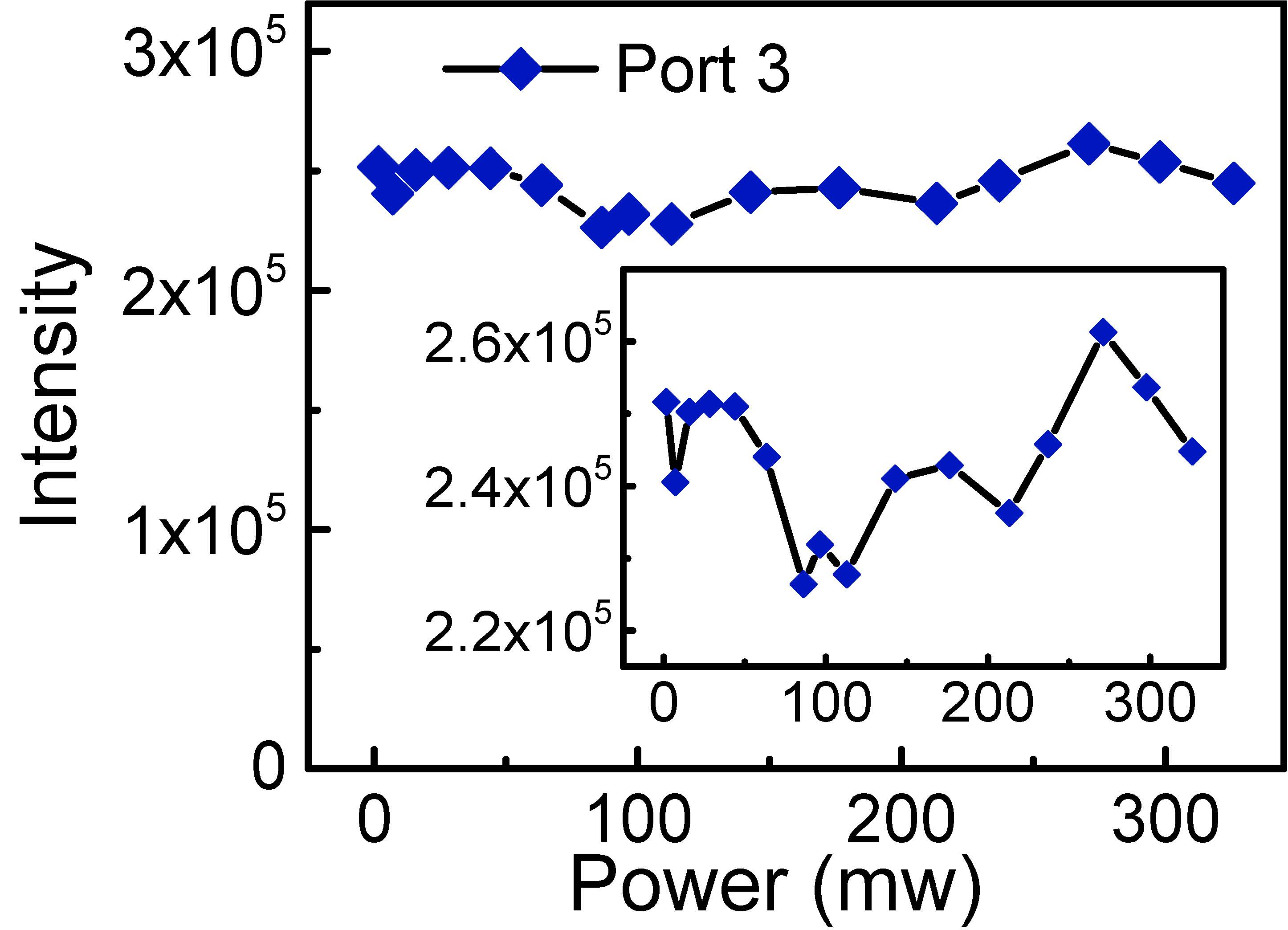}
\caption{\textbf{Intensity at output port 3 during the characterization of VS junction 2.} Output port 3 is disconnected to VS junction 2 while connected to VS junction 1. The intensity is measured when VS junction 2 is characterized and VS junction 1 undergoes a power cut. The irregular and small intensity fluctuation indicates the negligible thermal crosstalk.}
\label{figs9}
\end{figure*}

In addition, the negligibility of the thermal crosstalk can be further investigated by a comparison of the computing results of the photonic processor under different conditions. When Entry 2 serves as the input port and VS junction 2 is in balance mode, the photonic processor is programed to solve the SSP instance where $S=\{3, 5, 7, 11, 13, 17\}$. In theory, the working mode of VS junction 1 does not influence the computing results. We first apply zero current to VS junction 1. In this case, the power dissipation of the phase shifter in VS junction 1 is zero, making it impossible to introduce any thermal crosstalk. Therefore, the computing results are of high reliability, as demonstrated in Figs. S10a and S10b. As a contrast, we deliberately set VS junction 1 to balance mode (the dissipated power is 113.6 mw) and keep other settings of the photonic processor unchanged. The comparison of the computing results in the two cases (i.e., zero and nonzero dissipated power) enables us to investigate whether there is thermal crosstalk. As presented in Figs. S10c and S10d, the experimental results are highly similar to the case where no current is applied to VS junction 1, confirming the negligibility of the thermal crosstalk between the variable split junctions.

\begin{figure*}[htp]
\centering
\includegraphics[width=1\columnwidth]{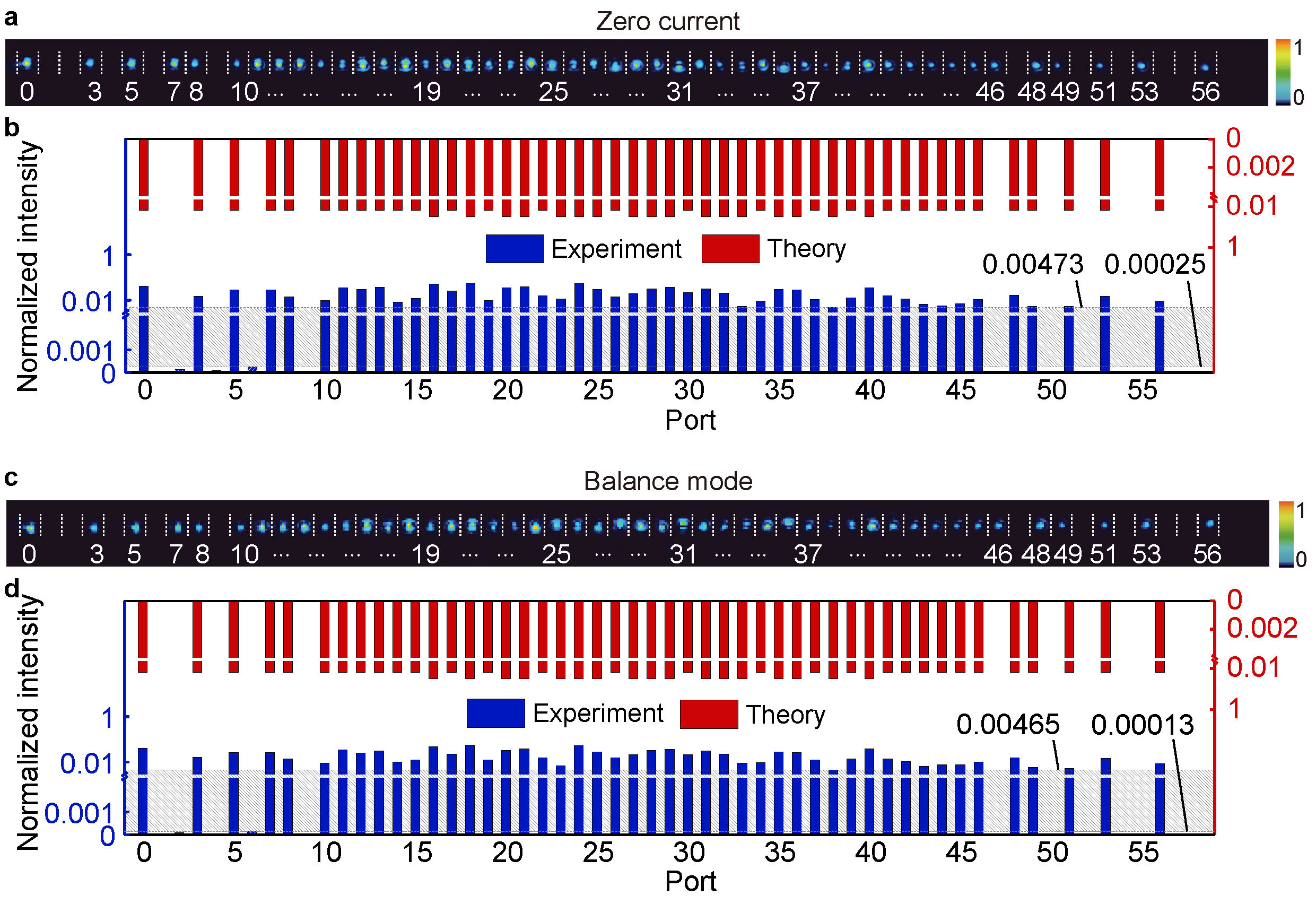}
\caption{\textbf{The computing results of the case \{3, 5, 7, 11, 13, 17\} under different conditions. } \textbf{a}, The experimental read-out and \textbf{b}, intensity distribution when there is no current applied to VS junction 1. \textbf{c}, The experimental read-out and \textbf{d}, intensity distribution when VS junction 1 is set to balance mode.}
\label{figs10}
\end{figure*}

\end{document}